\documentclass[pra,twocolumn,notitlepage]{revtex4}
\usepackage[utf8]{inputenc}
\usepackage{graphicx}
\usepackage{bm}
\usepackage{amsmath}

\usepackage{color}
\newcommand{\etal}{{\em et al. }}

\begin{document}

\title{Correspondence on ``Single-shot simulations of dynamic quantum many-body systems''}

\author{M.~K. Olsen$^{1}$, J.~F. Corney$^{1}$, R.~J. Lewis-Swan$^{2}$, and A.~S. Bradley$^{3}$}

%-----------------------------------------------------------------------
%\date{\today}
%------------------------------------------------------------------------    

\maketitle

{\it $^{1}$School of Mathematics and Physics, University of Queensland, Brisbane, QLD 4072, Australia. \\
$^{2}$Department of Physics, University of Colorado, Boulder, Colorado 80309-0440, USA. \\
$^{3}$Department of Physics, Centre for Quantum Science, and Dodd-Walls Centre for Photonic and Quantum Technologies, University of Otago, Dunedin, New Zealand.}

In a recent letter to Nature Physics, Sakmann and Kasevich claim to solve the many-body time dependent Schrödinger equation to simulate single experimental runs of interacting quantum systems~\cite{Kasevich}. In a subsequent comment, Drummond and Brand pointed out that the authors had, in two of their three results, only solved a truncated version of the full equation~\cite{PDDJB}. Further to these criticisms, we add some further comments particularly in reference to the truncated Wigner method (TWM) as applied to condensed Bose gases~\cite{Steeletal}.

The authors claim that the MCTDHB (multiconfigurational time-dependent Hartree method for bosons) method can perform single-shot simulations of many-body quantum experiments, whereas the TWM cannot because it is restricted to positive distributions. However, this argument overlooks the point that the implementation of the MCTDHB as described in the paper is also not exact. For example, the first two results in \cite{Kasevich} consider only two or four terms in expansions which formally require N terms for an N-body system.   The fact that the authors managed to reproduce the exact results for 10 atoms says little about the accuracy of their basis truncation for either 100 or 10,000 atoms.

The issue of non-Gaussian pure states and positive Wigner functions raised in~\cite{Kasevich} has already been discussed in\ \cite{JFCWig}.
Essentially, the TWM provides a joint probability distribution that is often a good approximation to the true Wigner distribution, and generally becomes more accurate for larger N.  It can give an accurate calculation of the moments for unitary evolution even when the true Wigner function is not positive definite~\cite{JFCWig}, and achieves this without introducing mixedness.

In so far as the predictions of the TWM are accurate, the joint probability distribution provides a hidden-variable 
explanation for symmetrically ordered observables.  It is then consistent to assign an element of reality to individual trajectories.  For other types of observables, the interpretation is not so straightforward, but the corrections to calculated averages are simple. For large enough N these contributions become negligible.

The interpretation of single trajectory TWM (or classical field method) simulations as possible experimental outcomes has succesfully described vortex formation in trapped condensates~\cite{Weiler}. Via appropriate noise in the initial conditions, it yields insight into interesting non-mean field physics captured 
in the deterministic evolution of a single trajectory~\cite{Norrie}.
Lewis-Swan \etal also show that TWM trajectories quantitatively reproduce the number distributions for a 
wide and well defined range of states~\cite{BobWig}. For a Bose-Hubbard model the results compare well with the exact values. 
These results support claims that individual trajectories have useful and practical relationship to experiment. 

The arguments provided by Sakmann and Kasevich concern the actual Wigner function, and are thus not relevant to the positive approximation to this sampled by the TWM.  If the TWM describes a particular physical scenario, i.e.\ the operator moments are accurate, then assigning physical significance to the quadrature values of individual trajectories requires no further approximation.   

In summary, MCTDHB and TWM are powerful methods with different limitations.  Their domains of applicability are complimentary, with MCTDHB demonstrably working well for small $N$, whereas TWM improves as $N$ increases.   Although care must be taken in the interpretation of single trajectories of TWM, their use in simulating experimental shots has some justification and has moreover proven insightful in understanding the physics of a number of experimental situations.

The authors declare that they have no competing financial interests.

\end{document}